\documentclass[conference]{IEEEtran}
\IEEEoverridecommandlockouts
\usepackage{cite}
\usepackage{amsmath,amssymb,amsfonts}
\usepackage{graphicx}
\usepackage{textcomp}
\usepackage{xcolor}
\usepackage{tikz}
\usetikzlibrary{quantikz}
\usepackage[frozencache,cachedir=.]{minted}
\usepackage{caption}
\usepackage{subcaption}
\usepackage{hyperref}
\usepackage{algorithm}
\usepackage{algcompatible}

\def\BibTeX{{\rm B\kern-.05em{\sc i\kern-.025em b}\kern-.08em
    T\kern-.1667em\lower.7ex\hbox{E}\kern-.125emX}}
\begin{document}

\title{Optimising Iteration Scheduling for Full-State Vector Simulation of Quantum Circuits on FPGAs
}

\author{\IEEEauthorblockN{Youssef Moawad}
\IEEEauthorblockA{\textit{School of Computing Science} \\
\textit{University of Glagow, UK}\\}
\and
\IEEEauthorblockN{Andrew Brown}
\IEEEauthorblockA{\textit{School of Mathematics} \\
\textit{University of Glagow, UK}\\}
\and
\IEEEauthorblockN{Ren\'e Steijl}
\IEEEauthorblockA{\textit{School of Engineering} \\
\textit{University of Glagow, UK}\\}
\and
\IEEEauthorblockN{Wim Vanderbauwhede}
\IEEEauthorblockA{\textit{School of Computing Science} \\
\textit{University of Glagow, UK}\\}
}

\maketitle

\begin{abstract}
As the field of quantum computing grows, novel algorithms which take advantage of quantum phenomena need to be developed. As we are currently in the NISQ (noisy intermediate scale quantum) era, quantum algorithm researchers cannot reliably test their algorithms on real quantum hardware, which is still too limited. Instead, quantum computing simulators on classical computing systems are used. In the quantum circuit model, quantum bits (qubits) are operated on by quantum gates. A quantum circuit is a sequence of such quantum gates operating on some number of qubits. A quantum gate applied to a qubit can be controlled by other qubits in the circuit. This applies the gate only to the states which satisfy the required control qubit state.

We particularly target FPGAs as our main simulation platform, as these offer potential energy savings when compared to running simulations on CPUs/GPUs. 

We focus on full-state-vector simulation of the quantum circuit model, which involves storing all possible states in memory  and applying each gate in sequence. Because every qubit can be in 2 states, the memory requirement scales exponentially with the number of qubits. In general, each quantum gate operation involves accessing the entire state vector in pairs and so the number of  steps to execute each gate operation grows exponentially with the number of qubits. Each added control to a gate halves the number of pairs that need to be accessed. 

In this work, we present a memory access pattern to optimise the number of iterations that need to be scheduled to execute a quantum gate such that only the iterations which access the required pairs (determined according to the control qubits imposed on the gate) are scheduled. We show that this approach results in a significant reduction in the time required to simulate a gate for each added control qubit. We also show that this approach  benefits the simulation time on FPGAs more than CPUs and GPUs and allows to outperform both CPU and GPU platforms in terms of energy efficiency, which is the main factor for scalability of the simulations.
\end{abstract}


\section{Introduction}

Quantum computing takes advantage of uniquely quantum phenomena like superposition and entanglement to allow for a new model of computation that can be up to exponentially more powerful than the classical model of computation, for specific problems. The problems to which quantum computing can be applied with a significant speedup remain limited and so research into further quantum algorithms is necessary.

Quantum computing simulation currently plays a key role in the development and verification of novel algorithms that can take advantage of the computational power offered by quantum computing.

In quantum computing, quantum bits (or \textbf{qubits}) take the place of classical bits as the units of information that are manipulated to perform computation. A qubit is described by two complex probability amplitudes:

\begin{equation}
\ket{q} = a\ket{0} + b\ket{1}  \mbox{\hskip 0.1cm ; \hskip 0.1cm}
|a|^2+|b|^2 = 1
\label{eq_qubit_def}
\end{equation}

Every qubit added to the system doubles the number of states that can exist in superposition. And so, in general, we need $2^n$ complex number to represent an $n$-qubit quantum system in a coherent state.

\subsection{Quantum Circuit Model}

\begin{figure}[!h]
    \centering
\begin{tikzpicture}
\node[scale=0.7]{
\begin{quantikz}
\lstick{$\ket{q1}$} & \ctrl{1} & \qw \rstick{$\ket{q1}$} \\
\lstick{$\ket{q0}$} & \targ &\qw & \qw \rstick{$\ket{q1}\oplus\ket{q0}$}
\end{quantikz}
\mbox{\hskip 0.80cm}
\begin{quantikz}
\lstick{$\ket{q1}$} & \octrl{1} & \qw \\
\lstick{$\ket{q0}$} & \gate{G} & \qw
\end{quantikz}
\mbox{\vspace{-0.5cm}\hskip 0.25cm$\equiv$\hskip 0.25cm}
\begin{quantikz}
\lstick{$\ket{q1}$} & \targ &\qw & \ctrl{1} & \targ &\qw &\qw \\
\lstick{$\ket{q0}$} & \qw &\gate{G} &\qw &\qw
\end{quantikz}
};
\end{tikzpicture}
\caption{Illustration of simple two-qubit quantum circuits.}
\label{fig_CircuitExample}
\end{figure}
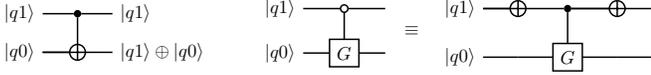

The most widely used model of quantum computation is the quantum circuit model, where algorithms are represented by sequences of quantum gate operations acting on one or more qubits, forming quantum circuits. Qubits are represented by horizontal lines in quantum circuit diagrams, stacked vertically for multi-qubit registers. Quantum gates are placed on these lines in square boxes. The $X$ gate ($X \equiv \begin{bmatrix} 0 & 1 \\ 1 & 0 \end{bmatrix}$) has the effect of flipping a qubit in the computational basis, such that $X\ket{0}=\ket{1}$, $X\ket{1}=\ket{0}$, by swapping of the qubit's probability amplitudes. The $X$ gate is often indicated by an xor symbol ($\oplus$) on a quantum circuit. This is the analogue of the classical NOT gate. The Hadamard gate ($H \equiv \frac1{\sqrt{2}}\begin{bmatrix} 1 & 1 \\ 1 & -1 \end{bmatrix}$) introduces a superposition on a qubit that is in the computational basis: $H\ket{0}=\frac1{\sqrt{2}}(\ket{0}+\ket{1})$ and $H\ket{1}=\frac1{\sqrt{2}}(\ket{0}-\ket{1})$. Controls can be imposed on gate application to entangle two or more qubits. A control qubit is indicated by a black dot on the qubit's line vertically connected to the gate. The left part of Figure \ref{fig_CircuitExample} demonstrates a controlled $X$ gate.

In more complex circuits, a series of such quantum gate operations is performed step-by-step. For illustration, the right-hand side of Figure \ref{fig_CircuitExample} shows how a general single-qubit gate $G$ with a negative control (here $\ket{q1}=\ket{0}$) can be implemented using a sequence of three gate operations. The number of time slices required to run the circuit (i.e. circuit slices with gates that affect mutually exclusive qubits) defines the \textit{circuit depth}.

\section{Related Work}

Extensive research has resulted in a range of highly-optimized quantum computer simulators for multi-core and distributed computing architectures. These include Intel Quantum Simulator \cite{smelyanskiy_qhipster_2016}, ProjectQ \cite{steiger_projectq_2018}, JUMPIQCS \cite{de_raedt_massively_2007}, Qrack \cite{strano_qrack_nodate}, and QCGPU \cite{kelly_simulating_2018}.

The simulation of quantum computers and, specifically, quantum-circuit implementations of algorithms on FPGAs has been the topic of more recent research works. Examples of works focusing on FPGAs include\cite{khalid_fpga_2004}, \cite{aminian_fpga-based_2008}, \cite{conceicao_efficient_2015}, \cite{lee_fpga-based_2016}, \cite{mahmud_scalable_2018}, \cite{pilch_fpga-based_2019}, \cite{mahmud_efficient_2020}, \cite{khalid_fpga_2021}, and \cite{bonny_emulation_2020}.

A fuller summary on recent works on cluster-based and FPGA-based simulation of quantum circuits is presented in the authors' earlier work \cite{moawad_investigating_2022}.

\section{Quantum Circuit Simulation}

In this work we target full state vector simulation as it is the most general simulation approach, allowing for complete control and introspection of the state of the system at any point during the computation. 

In our simulator, we support single qubit gates with arbitrary number of controls. Our supported gate set is currently comprised of the Hadamard gate ($H$), the Pauli gates ($X$, $Z$, and $Y$) and phase rotation gates ($R_m$). We allow an arbitrary number of controls on any gate, up to a compile-time architecture parameter for the maximum possible number of controls per gate. This set of gates supports universal quantum computation.

For an n-qubit simulation, $2^n$ complex numbers need to be stored in memory. In general, to simulate the operation of one quantum gate, the entire memory space needs to be accessed. The memory space is accessed in pairs, and the access pattern depends on the gate's target qubit index, $t$, where $t=0$ represents the least significant qubit in the register. In particular, the strides between the elements of each pair is given by $2^t$. Consider the application of a general gate, $G = \begin{bmatrix} a & b \\ c & d \end{bmatrix}$, where $a, b, c, d \in \mathbb{C}$, on a qubit $t$. If the initial quantum state is defined by $\ket{\Psi}_0$ and the state after applying the gate by $\ket{\Psi}_1$ then:
\begin{equation}
\ket{\Psi}_1 = \sum_{k=0}^{2^n} C_k^{(1)} \ket{k}
\label{eq_psi1_state}
\end{equation}
where complex amplitudes $C_k^{(1)}$ are related to previous amplitudes $C_k^{(0)}$ as:
\begin{eqnarray}
C_k^{(1)} &=& a C_k^{(0)} + b C_{k+2^t}^{(0)}
\label{eq_psi1_G_amplitudes}\\
C_{k+2^t}^{(1)} &=& c C_k^{(0)} + d C_{k+2^t}^{(0)}\nonumber
\end{eqnarray}
for $k\in[0, 2^{n-1}-1]$. 

Each quantum gate requires $2^{n-1}$ iterations to be simulated, where each iteration accesses a unique pair of complex numbers in the memory. Because of this dependency, concurrent simulation of gate operations is not possible. We can however parallelise the iterations of any one gate.

\begin{figure}
    \centering
    \includegraphics[scale=0.1]{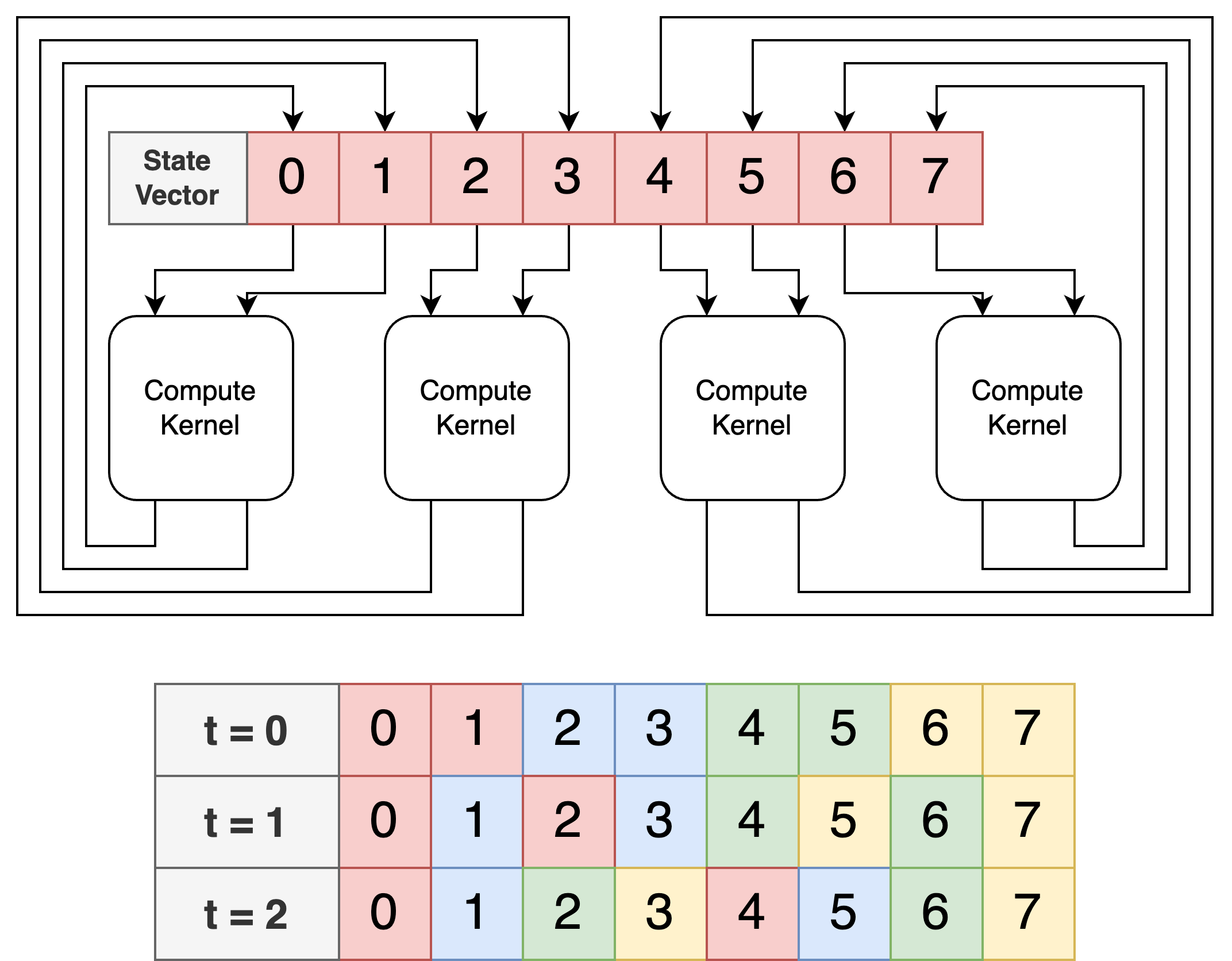}
    \caption{Memory access pattern example for 3-qubit full state vector simulation. The numbers on the state vector represent the index of the complex probability amplitude. The box in the bottom part of the figure shows how the amplitudes are accessed depending on different target qubits. Like-coloured boxes are accessed and computed on together.}
    \label{fig:qwm_memory_access}
\end{figure}

Figure \ref{fig:qwm_memory_access} shows the access pattern for the execution of a non-controlled gate on an example 3-qubit register. The top half of the figure shows the example for $t=0$, where the pairs are constituted by contiguous elements from the state vector. The compute units update the pairs based on equation \ref{eq_psi1_G_amplitudes}, and stores them back into the state vector memory. The bottom part of the figure shows the access pattern for higher values of $t$.

When controls are imposed on a gate, \textbf{each control halves the set of pairs that need to be updated in memory}. The condition to apply a controlled gate to an iteration pair is that the value of the bit (of the amplitude's index in the state vector in binary) in the same position as the control qubit in the register is 1. It can be observed that since the two elements constituting a pair will only differ in the target qubit, then the controls will cancel whole iterations at a time (as opposed to possibly updating a single element of the pair); i.e. pairs are either entirely updated together or not at all. 

Traditionally, this is handled by a check in the execution of an iteration, which determines whether the memory access should go ahead after the pair indices have been computed based on the target qubit's index. On a CPU/GPU with dynamically scheduled iterations, this does not cause a loss in performance as no cycles are scheduled for reading and writing to memory until the control flow determines that this should happen. On an FPGA, however, the potential memory accesses are statically scheduled and so clock cycles are always allocated for them; wasting cycles when the control flow determines that no computation should take place.

\subsection{Implementation}

We use OpenCL kernels inspired by \cite{kelly_simulating_2018} to facilitate the simulation of a quantum gate. Figure \ref{alg:sim_baseline_kernel} demonstrates pseudocode for the execution of an iteration. The kernels are built for the FPGA board using Intel's OpenCL FPGA SDK, as NDRange kernels. The global IDs for the NDRange kernels are the iteration indices and so go from $0$ to $2^{n-1}-1$ for $n$ qubits. The host schedules an NDRange kernel in this way for every gate in the circuit, serially. We use 2 32-bit floating point numbers to represent each complex amplitude in memory.

\begin{figure}
{\small
\begin{minted}{c}
/* The following are provided by 
   the host as kernel parameters */
uint t = Gate target qubit
uint c0,c1,...,c_n_c = Gate control qubits
cfloat mat0,mat1,mat2,mat3 = Gate matrix
cfloat *state = global memory pointer
                
uint i = NDRange global ID, corresponding
        to the iteration index

// Compute pair element indices
uint pairElement1 = ithCleared(i,t)
uint pairElement2 = pairElement1 + 2^t

// Check controls
bool perform = true
for(c in controls) 
  perform &= (pairElement1 satisfies c)
/* If any of the controls are not satisfied
   perform will be false */

if(perform) {
  /* Update the state in memory by 
  multiplication with the gate matrix */
  cfloat2 inVec = 
    state[pairElement1, pairElement2]
  state[pairElement1, pairElement2] = 
    [mat0 * inVec[0] + mat1 * inVec[1],
     mat2 * inVec[0] + mat3 * inVec[1]]
} else return from the kernel
\end{minted}
}
\caption{Full state vector gate simulation kernel pseudocode. $2^{n-1}$ iterations of this kernel are scheduled by default. It can be seen that the memory access depends on dynamic control flow based on the control qubits imposed on the quantum gate.} \label{alg:sim_baseline_kernel}
\end{figure}

\subsection{Optimising controls processing}

The primary contribution of this work comes from realising that we can schedule exactly as many iterations which perform memory accesses as are required for the gate, taking the gate's controls into account. To demonstrate this, we introduce the concept of an iteration index set in the context of full state vector simulation. As described above, to simulate a quantum gate over an $n$-qubit register, $2^{n-1}$ iterations are required, giving an iteration index set of $[0, 2^{n-1}-1]$. Define this set as $I_g$ for global iteration index set. These are the indices which can be plugged into the \mintinline{c}{ithCleared(i,t)} function \cite{kelly_simulating_2018} along with the target qubit $t$ to return the index of the first element in the pair of amplitudes that need to be processed for any given iteration $i \in I_g$.

Our goal is to be able to schedule only the number of iterations that are required taking into account each added control, i.e. introduce a \textbf{reduced iteration index set} $I_r = [0, 2^{n-n_c-1}-1]$, where $n_c$ is the number of controls of the gate. 

The challenge is to map this reduced set $I_r$ back to the global set $I_g$, as directly scheduling $I_r$ would result in incorrect calculations. The idea is to iteratively map the smaller iteration sets back to $I_g$, considering each control qubit, selecting the values of $i$ from $I_g$ (which can be plugged into \mintinline{c}{ithCleared}) that correspond to $I_r$ taking into account the set of controls applied to the gate $C = \{c_0, c_1, ..., c_{n_c-1}\}$. 

This means we need a map from the values of $I_r$ to the values of $I_g$. For a single control, let $I_{r_0}$ be the reduced iteration set; this will have half the cardinality of $I_g$. If we introduce a further control, let the corresponding reduced iteration set be $I_{r_1}$ (which will have half the cardinality of $I_{r_0}$); as long as this further control is higher in the register than the first control, we can map from $I_{r_1}$ to $I_{r_0}$, and then finally map from $I_{r_0}$ to $I_g$. This is the general idea for handling higher numbers of controls: iteratively map from the smaller iteration sets until the global iteration index in $I_g$ is reached.

To realise this, we map out the required memory accesses for differently controlled gates to find a pattern. We start from the access pattern demonstrated in Figure \ref{fig:qwm_memory_access} and rearrange the amplitudes such that iteration pairs are contiguous (they are not actually rearranged in memory, rather this is just for demonstration). We then impose controls, in an ascending order, for every target qubit example. We make an exhaustive list of controls: for a 4-qubit example, there will be three cases with 1 control qubit, three cases with 2 control qubits, and one case with 3 control qubits. When controls are arranged in a logical order across different target examples, a pattern emerges in the iterations which are skipped. 

\begin{figure}
    \centering
    \includegraphics[scale=0.1]{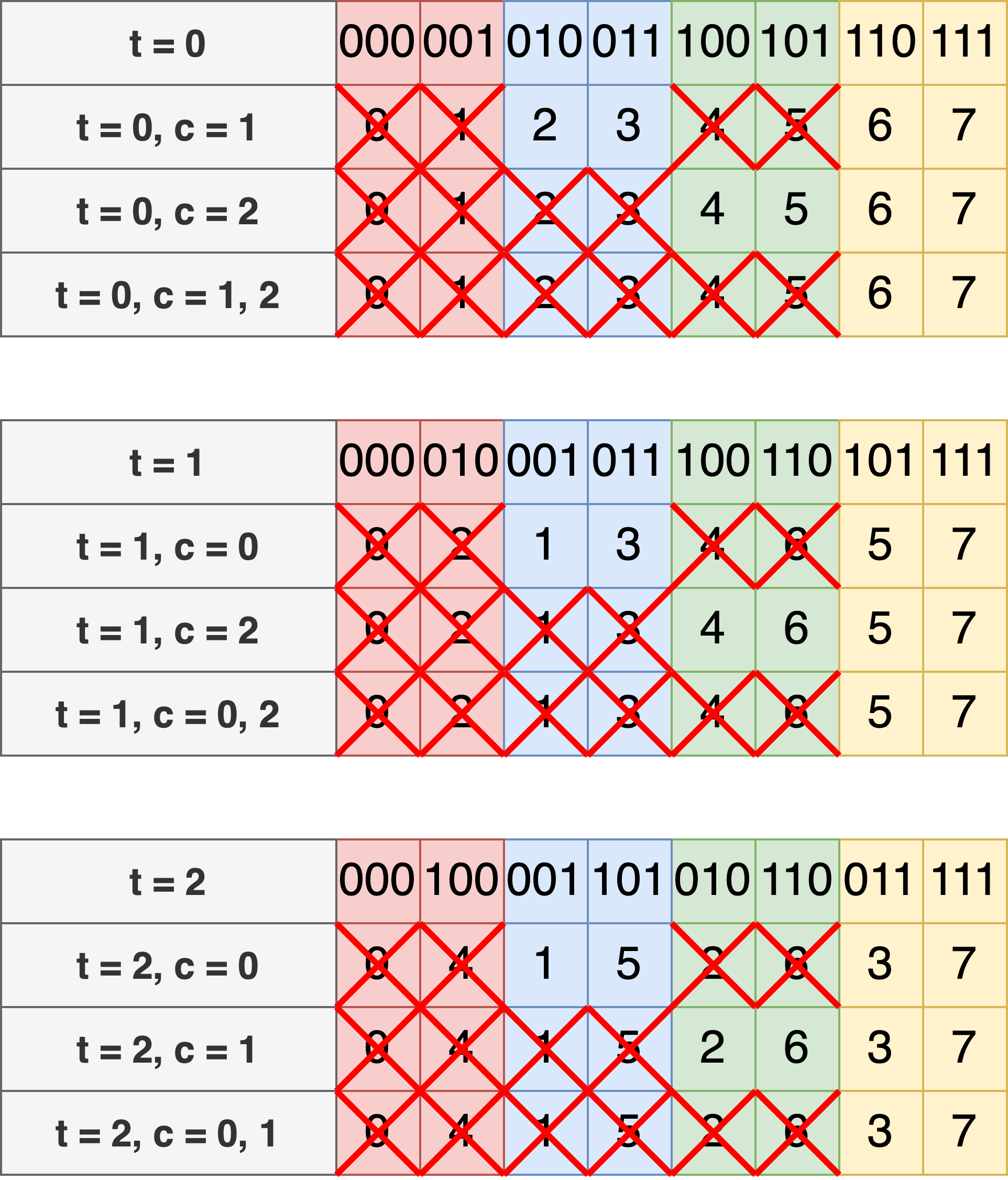}
    \caption{Access patterns for controlled gate applications for a 3-qubit register. For demonstration, the boxes representing the amplitudes are rearranged such that required pairs are contiguous, in contrast to Figure \ref{fig:qwm_memory_access}. Crossed out pairs indicate skipped iterations due to the controls imposed on the gate. It can be observed that the pattern of control skips is the same when controls are arranged in ascending order, as shown.}
    \label{fig:qwm_controls_memory_access}
\end{figure}

For brevity, Figure \ref{fig:qwm_controls_memory_access} shows the example for a 3-qubit register. Note that the first row in each target example (representing the uncontrolled case) is now shown in binary, to make it easier to recognise the control condition for the controlled cases, and the whole table is rearranged such that iteration pair elements are contiguous. For each controlled case, we  cross out the iteration pairs which do not satisfy the controls (where the control qubits are 0 in the binary representation). Regardless of the target qubit, the same \textbf{iteration skips} pattern emerges.

In order to encode these iteration skips, we start by introducing the concept of an \textbf{adjusted control}, which is a re-indexing of the control qubits relative to the target qubit: if the control qubit is greater than the target qubit, subtract one, otherwise keep its original value. This is shown in Eq. \ref{eq:adj_controls_formula}. This method gives the same values of adjusted controls for all the \textbf{control qubit enumerations} for different values of the target qubit; e.g. for the 3-qubit register demonstrated in Figure \ref{fig:qwm_controls_memory_access}, the method gives us adjusted control values of $c_{adj} = 0$, $c_{adj} = 1$, and $c_{adj} = 0,1$ for the three possible control qubit cases. We can then compute a \textbf{skip interval} corresponding to each adjusted control value as $2^{c_{adj}}$. Based on the computed skip interval, we can map any iteration index belonging to a reduced iteration index set ($i_{r_{k+1}}$) to a higher iteration index ($i_{r_k}$) by adding to it, following the iterative formula shown in Equation \ref{eq:controls_iterative_formula} where we can consider $i_g$, the global iteration index, as $i_{r_{-1}}$.

\begin{equation}
c_{adj} = 
\left\{ 
  \begin{array}{ c l }
    c - 1 & \quad \textrm{if } c > t\\
    c     & \quad \textrm{otherwise}
  \end{array}
\right.
\label{eq:adj_controls_formula}
\end{equation}

\begin{equation}
i_{r_k} = i_{r_{k+1}} + (\lfloor i_{r_{k+1}} / 2^{c_{adj}} \rfloor + 1) \times 2^{c_{adj}}
\label{eq:controls_iterative_formula}
\end{equation}

To illustrate this, we take the $t=1$ case for a 3-qubit register as an example (the middle table in Figure \ref{fig:qwm_controls_memory_access}). For 3 qubits, the global iteration index set is $I_g = \{0,1,2,3\}$ (as there are $2^3=8$ amplitudes, the number of iterations is half the size of the state vector). When one control is imposed ($c_0 = 0$ or $c_0 = 2$), the reduced iteration index set is $I_{r_0} = \{0,1\}$, and these are the values which the NDRange kernel will be scheduled with (i.e. the values that \mintinline{C}{get_global_id()} will return). For $c_0 = 0$, the adjusted control remains the same as $c_0 < t$, so $c_{0_{adj}} = 0$. The skip interval is then computed as $2^{c_{0_{adj}}} = 2^0 = 1$. 

Then, following Equation 4, $I_{r_0}$ can be mapped to $I_g$ in this way: $\{0,1\} \rightarrow \{0+(0/1+1)\times1, 1+(1/1+1)\times1\} = \{1,3\}$, meaning the second and fourth global iterations are the required ones, which matches the result in the figure. For the case where $c_0 = 2$, $c_0 > t$ and so the adjusted control is $c_{0_{adj}} = c_0 - 1 = 1$, giving a skip interval of $2^1 = 2$. Following the same formula as the previous case, we can map the reduced iteration index set to the global one in this way: $\{0,1\} \rightarrow \{0+(0/2+1)\times2, 1+(\lfloor1/2\rfloor+1)\times2\} = \{2,3\}$, meaning the last two global iterations are the desired ones, which again can is verified by the figure. Finally, for the case where we have two controls $c_0 = 0, c_1 = 2$, we perform two mappings: one from $I_{r_1} = \{0\}$ (corresponding to the two imposed controls) to $I_{r_0} = \{0,1\}$ (corresponding to one imposed control) and then to $I_{r_{-1}} = I_g = \{0,1,2,3\}$, the global iteration index set. 

To go from $I_{r_1}$ to $I_{r_0}$, we take the first control $c_0 = 0$ and apply the formula: $c_{0_{adj}} = c_0 = 0$ as $c_0 < t$, and the skip interval is again $2^0 = 1$. The single element, $i_{r_1} = 0$ in $I_{r_1}$ is then mapped to $i_{r_0} = 0+(0/1+1)\times1 = 1$ in $I_{r_1}$. We then perform the second mapping, for $c_1 = 2$: $c_{1_{adj}} = c_1 - 1 = 1$ as $c_1 > t$, and the skip interval is $2^1 = 2$. Applying the formula gives $i_g = 1+(\lfloor1/2\rfloor+1)\times2 = 3$, meaning only the final iteration in the global iteration set is to be performed. The only condition for this formula to function correctly is that the controls have to be in ascending order, i.e. strictly $c_0 < c_1 < ... < c_{n_c-1}$.

With this formula, we can schedule the compute kernels with exactly the number of iterations that will always update the memory. We use the formula to iteratively go from a reduced iteration index to the equivalent global iteration index allowing us to use the previously used memory indexing strategy based on the \mintinline{C}{ithCleared} function. Figure \ref{code:opencl_formula_impl} shows the implementation of the formula in the OpenCL kernel, for up to 2 controls. This block would go above the pair element index computation in the pseudocode for the baseline kernel shown above in Figure \ref{alg:sim_baseline_kernel}, and then $i_g$ would be used in the computation of \mintinline{c}{pairElement1} in place of $i$ in the \mintinline{c}{ithCleared} function. The controls check block that utilises a boolean flag \mintinline{c}{perform} would no longer be needed and there would be no control flow checks on the memory access.

\begin{figure}
    \centering
    {\small
    \begin{minted}{C}
    uint i_g = get_global_id(0);
    
    // adjusted control:
    uint c_adj0 = c0 > t ? c0-1 : c0; 
    // skip interval:
    uint sInt0 = 1 << c_adj0;
    i_g += ((i_g/sInt0) + 1) * sInt0;
    
    uint c_adj1 = c1 > t ? c1-1 : c1;
    uint sInt1 = 1 << c_adj1;
    i_g += ((i_g/sInt1) + 1) * sInt1;
    \end{minted}
    }
    \caption{Implementation of the iterative formula for going from a reduced iteration index set to the global set in the OpenCL kernel. Example shown for up to 2 controls.}
    \label{code:opencl_formula_impl}
\end{figure}

\section{Evaluation}

We evaluate our optimised iteration scheduling by building two target architectures: a baseline based on the kernel pseudocode in Figure \ref{alg:sim_baseline_kernel} which always schedules $2^{n-1}$ iterations, and an optimised iterations architecture scheduling $2^{n-n_c-1}$ iterations, where $n_c$ is the number of controls imposed on the gate being scheduled. 

We utilised a system with a dual Intel Xeon E5-2609 V2 2.5 GHz processor and 64 GB DDR3 1.6 GHz RAM running Scientific Linux 6.8. Our system hosts a single Nallatech PCIe-385N A7 FPGA board with 8 GB DDR3 RAM connected through a PCIe 2 connection and we used the Intel SDK for OpenCL version 17.1 to compile for and program the board. The host code is compiled using G++ (GCC version 4.7.2). We use the same FPGA host system to evaluate the CPU and we run the same OpenCL kernel with the same host code. To evaluate on the GPU, we have access to a different host with an NVIDIA GK110B (GeForce GTX TITAN Black), with access to 6GB of GDDR5  VRAM. The GPU system compiles the C++ host code with G++ (GCC version 5.4.0).

In order to be competitive with distributed computing simulation methods, it will be necessary to move to clusters of FPGAs. In the case of supercomputing clusters, energy consumed becomes an important metric, and so we choose to focus on the energy consumed by a device to simulate an evaluation circuit. Our FPGA's power consumption is rated at 25W \cite{segal_high_2014}, while our CPU consumes 160W and the GPU consumes 250W. To compute the energy consumed during the execution of a circuit, we multiply the total time of the circuit by the target platform's power consumption.

\begin{table}[h]
\centering 
\begin{tabular}{c|c|c|c}
  Kernel          & Baseline        & Optimised iterations  & Total Available \\
  \hline
  ALUTs           & $59772 (17\%)$  & $60029 (17\%)$        & $345200$ \\
  FFs             & $71743 (10\%)$  & $71941 (10\%)$        & $690400$ \\
  RAMs            & $408 (20\%)$    & $408 (20\%)$          & $2014$ \\
  DSPs            & $16 (1\%)$      & $16 (1\%)$            & $1590$ \\ 
  $F_{max}$ (MHz) & $296.02$        & $308.35$              & $-$ \\
\end{tabular}
\caption{FPGA resource utilisation and maximum frequency for the baseline and optimised iterations kernels with 2 maximum controls per gate. It should be noted that changing the maximum number of controls allowed per gate (which is a compile time parameter) changes these figures minimally.}
\label{tab_FPGAResourceUtil}
\end{table}

For the purpose of evaluating the optimised iteration scheduling technique, we choose to synthesise both target architectures with one compute unit on the FPGA. As such we also compare against the CPU and GPU running with one OpenCL compute unit. Table \ref{tab_FPGAResourceUtil} shows the resource utilisation on the FPGA for both architectures with a maximum of 2 controls per gate. We observe that both architectures use roughly the same amount of resources and the optimised iterations architecture is able to run at a slightly higher maximum frequency.

We choose three algorithms as evaluation targets: the Quantum Fourier Transform (QFT), which forms an important part of several other quantum algorithms, an integer squaring circuit, and a streaming circuit.

\subsection{Quantum Fourier Transform}

\begin{figure}
    \centering
    \begin{tikzpicture}
    \node[scale=0.4]{
    \begin{quantikz}[row sep=0.3cm, column sep=0.15cm]
  \lstick{$q_0$} & \gate{H} & \gate[style={inner xsep=0pt}]{R_2} & \gate[style={inner xsep=0pt}]{R_3} & \gate[style={inner xsep=0pt}]{R_4} & \gate[style={inner xsep=0pt}]{R_5} & \qw & \qw & \qw & \qw & \qw & \qw & \qw & \qw & \qw & \qw & \qw & \qw \\
  \lstick{$q_1$} & \qw & \ctrl{-1} & \qw & \qw & \qw & \gate{H} & \gate[style={inner xsep=0pt}]{R_2} & \gate[style={inner xsep=0pt}]{R_3} & \gate[style={inner xsep=0pt}]{R_4} & \qw & \qw & \qw & \qw & \qw & \qw & \qw & \qw \\
  \lstick{$q_2$} & \qw & \qw & \ctrl{-2} & \qw & \qw & \qw & \ctrl{-1} & \qw & \qw & \gate{H} & \gate[style={inner xsep=0pt}]{R_2} & \gate[style={inner xsep=0pt}]{R_3} & \qw & \qw & \qw & \qw & \qw \\
  \lstick{$q_3$} & \qw & \qw & \qw & \ctrl{-3} & \qw & \qw & \qw & \ctrl{-2} & \qw & \qw & \ctrl{-1} & \qw & \gate{H} & \gate[style={inner xsep=0pt}]{R_2} & \qw & \qw & \qw \\
  \lstick{$q_4$} & \qw & \qw & \qw & \qw & \ctrl{-4} & \qw & \qw & \qw & \ctrl{-3} & \qw & \qw & \ctrl{-2} & \qw & \ctrl{-1} & \qw & \gate{H} & \qw \\
    \end{quantikz}
    };
    \end{tikzpicture}
    \caption{Example $5$-qubit QFT circuit.}
    \label{fig:qft_circuit}
\end{figure}
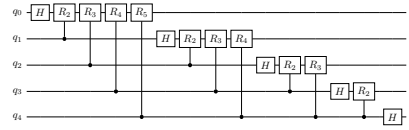

The Quantum Fourier Transform is a quantum algorithm that generalises the classical Fourier Transform to the quantum realm. It is a key component in several quantum algorithms, including Shor's algorithm \cite{shor_polynomial-time_1997}, QFT-based arithmetic algorithms \cite{draper_addition_2000}, and quantum phase estimation algorithms. The QFT maps an input quantum state to its Fourier transformed state in polynomial quantum time, allowing for the efficient manipulation of quantum information. The QFT circuit has a staircase pattern of Hadamard and controlled phase rotation gates. An example 5-qubit QFT circuit is shown in Figure \ref{fig:qft_circuit}; generalised n-qubit QFT circuits can be constructed following the same pattern.

\begin{figure}
    \centering
    \includegraphics[scale=0.275]{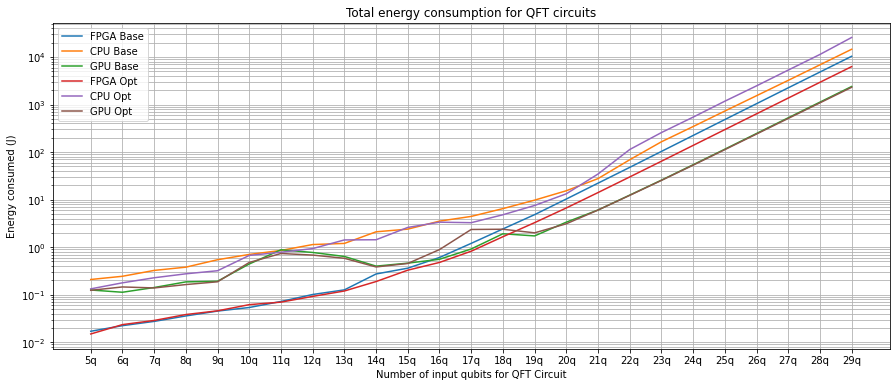}
    \caption{Total energy consumption of QFT circuits.}
    \label{fig:qft_energy}
\end{figure}

Figure \ref{fig:qft_energy} shows the total energy consumption for running QFT circuits of various circuit widths across the three evaluation platforms and both architectures.

\begin{table}[h]
\centering
\begin{tabular}{c|c|c}
  Platform and Kernel   & Total time (s)    & Total energy (J) \\
  \hline
  Baseline FPGA         & 413.83            & 10345.73 \\
  Optimised FPGA        & 251.00            & 6275.05 \\
  Baseline CPU          & 90.88             & 14540.74 \\
  Optimised CPU         & 160.93            & 25748.23 \\
  Baseline GPU          & 9.58              & 2395.93 \\
  Optimised GPU         & 9.22              & 2306.12 \\
\end{tabular}
\caption{QFT29 total time and energy summary.}
\label{tab:qft29_summary}
\end{table}

\begin{table}[h]
\centering
\begin{tabular}{c|c|c}
  Platform and Kernel   & Total time (s)    & Total energy (J) \\
  \hline
  Baseline FPGA         & 909.23            & 22730.74 \\
  Optimised FPGA        & 501.32            & 12533.00 \\
  Baseline CPU          & 174.71             & 27953.62 \\
  Optimised CPU         & 148.97            & 23834.92 \\
  Baseline GPU          & 16.52              & 4129.95 \\
  Optimised GPU         & 13.26              & 3314.17 \\
\end{tabular}
\caption{SQ9 (29 qubits) total time and energy summary.}
\label{tab:sq29_summary}
\end{table}

\begin{table}[h]
\centering
\begin{tabular}{c|c|c}
  Platform and Kernel   & Total time (s)    & Total energy (J) \\
  \hline
  Baseline FPGA         & 26.07            & 651.66 \\
  Optimised FPGA        & 3.78            & 94.44 \\
  Baseline CPU          & 10.36             & 1656.83 \\
  Optimised CPU         & 2.09            & 334.01 \\
  Baseline GPU          & 1.73              & 434.46 \\
  Optimised GPU         & 0.95              & 238.30 \\
\end{tabular}
\caption{29-qubit streaming total time and energy summary.}
\label{tab:stream29_summary}
\end{table}

\subsection{Squaring Circuits}

\begin{figure}
    \centering
    \includegraphics[width=8cm,height=4cm]{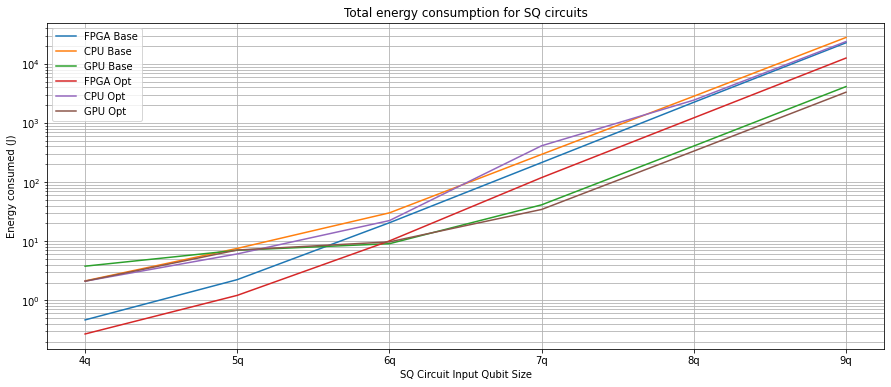}
    \caption{Total energy consumption of SQ circuits.}
    \label{fig:sq_energy}
\end{figure}

In \cite{moawad_investigating_2022}, the authors demonstrate squaring circuits that use controlled Cuccaro adders \cite{cuccaro_new_2004} and shift operators. We use these circuits as our second evaluation target. 
Cuccaro adders use quantum gates with up to two controls, and so our squaring circuits use gates with up to three controls, as they use controlled Cuccaro adders. Figure \ref{fig:sq_energy} shows the energy consumption of squaring circuits with different input bit sizes. An $n$-bit input squaring circuit uses $3\times n + 2$ qubits, and so our highest qubit example (the 9-bit input) uses 29 qubits.

\subsection{Streaming Circuits}

\begin{figure}
    \centering
\begin{tikzpicture}
\node[scale=0.6]{
\begin{quantikz}[row sep=0.3cm, column sep=0.15cm]
\lstick{$\ket{x0}$} &\targ{} &\ctrl{1} &\ctrl{1} &\ctrl{1}  &\qw &\qw \\
\lstick{$\ket{x1}$} &\qw &\targ{} &\ctrl{1} &\ctrl{1} &\qw &\qw \\
\lstick{$\ket{x2}$} &\qw &\qw &\targ{} &\ctrl{1} &\qw &\qw \\
\lstick{$\ket{x3}$} &\qw &\qw &\qw &\targ{} &\qw &\qw \\
\end{quantikz}
};
\end{tikzpicture}
\caption{4-qubit streaming circuit. A general $n$-qubit streaming circuit can be constructed in the same pattern, which will contain gates with up to $n-1$ controls.}
\label{fig:streaming_circ}
\end{figure}
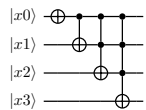

\begin{figure}
    \centering
    \includegraphics[scale=0.275]{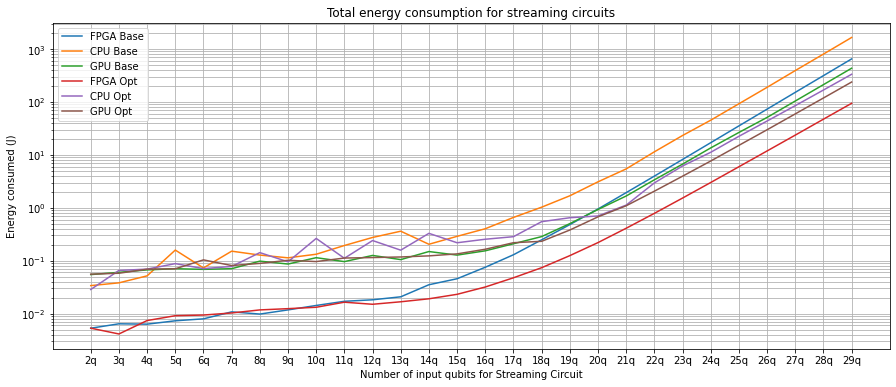}
    \caption{Total energy consumption of streaming circuits.}
    \label{fig:stream_energy}
\end{figure}

Our final evaluation target is a streaming circuit useful in quantum circuit implementations of the Lattice Boltzmann model, which was used in modelling quantum walks on a one- and two-dimensional lattices  \cite{todorova_quantum_2020}. An example 4-qubit streaming circuit is shown in \ref{fig:streaming_circ}. We can construct and run similar circuits with up to 29-qubits. The reason these circuits are chosen is because an $n$-qubit streaming circuit contains gates with up to $n-1$ controls, making it a good "best-case" test circuit for our optimisation. Figure \ref{fig:stream_energy} shows the energy consumed for different streaming circuits.

\subsection{Discussion}

As expected, circuits where more gates have high number of control qubits benefit most from our optimisation. Tables \ref{tab:qft29_summary}, \ref{tab:sq29_summary}, and \ref{tab:stream29_summary} present summaries of the timing and energy consumed for the largest circuits (all 29 qubits) for each of the evaluation algorithms. For the QFT (which has at most one control on a gate), we see almost a $2\times$ improvement for the FPGA, while the CPU performance worsens (a result of spending extra cycles computing the optimised indexing formula), and the GPU performance is mostly unaffected. For the squaring circuit (which has up to 2 more controls compared to the QFT), we see that all platforms benefit from utilising the optimisation, with the FPGA again benefiting the most at almost $2\times$ improvement in time and energy consumed. The streaming circuit, which has increasingly higher number of control qubits across its gates, demonstrates the most benefit for the FPGA simulation, at almost a $7\times$ improvement over the baseline. This testcase shows that it is possible for the FPGA simulation to outperform the CPU and the GPU in energy efficiency, even though our FPGA is overdimensioned for the circuit.

In addition, based on resource utilisation demonstrated in Table \ref{tab_FPGAResourceUtil}, we could run the architecture on up to 4$\times$ smaller FPGA, e.g. a Stratix V D3, which would result in better energy efficiency than the GPU-based simulation for all test cases.

\section{Conclusion}

In this work, we presented a technique for optimising the number of iterations that need to be scheduled for the execution of a quantum gate, ensuring that every iteration does useful work. We have demonstrated the specific advantage of our technique for FPGA-based simulation.

In the future, we intend to extend our FPGA architecture with support for multiple compute units to maximise resource utilisation. We aim to introduce further optimisations including gate fusion, fixed-point precision, and compute tables. 

\bibliographystyle{plain}
\bibliography{references}

\end{document}